\documentclass[journal=jacsat,manuscript=article]{achemso}

\usepackage[version=3]{mhchem} 
\usepackage{hyperref}
\usepackage{cleveref}
\usepackage {graphicx}
\usepackage{braket}
\usepackage{titlesec}
\usepackage{xcolor}
\usepackage{ulem}

\usepackage{array}
\usepackage{makecell}
\usepackage{diagbox}

\author{Keren Wang}
\altaffiliation{These authors contributed equally to this work}
\affiliation[Unknown University]
{College of Physics, Sichuan University, Chengdu 610064, China}

\author{Kaili Sun}
\altaffiliation{These authors contributed equally to this work}
\affiliation[Unknown University]
{Shandong Provincial Key Laboratory of Optics and Photonic Devices, Center of Light Manipulation and Applications, School of Physics and Electronics,
Shandong Normal University, Jinan 250358, China}

\author{Jing Du}
\affiliation[Unknown University]
{College of Physics, Sichuan University, Chengdu 610064, China}

\author{Peijuan Dai}
\affiliation[Unknown University]
{College of Physics, Sichuan University, Chengdu 610064, China}

\author{Hao Zhou}
\affiliation[Unknown University]
{College of Electronics and Information Engineering, Sichuan University, Chengdu 610064, China}

\author{Lujun Huang}
\email{ljhuang@phy.ecnu.edu.cn}
\affiliation[Unknown University]
{State Key Laboratory of Precision Spectroscopy, School of Physics and Electronic Science, East China Normal University, Shanghai 200241, China}

\author{Zhanghua Han}
\email{zhan@sdnu.edu.cn}
\affiliation[Unknown University]
{Shandong Provincial Key Laboratory of
Optics and Photonic Devices, Center of Light Manipulation and Applications, School of Physics and Electronics,
Shandong Normal University, Jinan 250358, China}

\author{Wei Wang}
\email{w.wang@scu.edu.cn}
\affiliation[Unknown University]
{College of Physics, Sichuan University, Chengdu 610064, China}

\title{Broadband-operational orbital angular momentum generation in nonlocal metasurfaces with maximum efficiency approaching 80\% }

\abbreviations{}

\begin{document}

\begin{abstract}

Nonlocal metasurfaces provide a compact route to generating momentum-space optical vortices but are limited by steep dispersion typically associated with high-quality (Q) factor resonances, resulting in narrowband and inefficient operation. Here, we introduce a reflection-type nonlocal metasurface that hybrid-couples a bound state in the continuum (BIC) with two degeneracy points (DPs). This engineered interaction enables on-demand control of dispersion, radiative Q-factors, and polarization states of guided resonances, yielding quasi-flat dispersion and enhanced scattering strength. Full-wave simulations predict near-unity on-resonance conversion and overall efficiencies above 90\%, representing a three- to fourfold efficiency improvement and more than fifteenfold bandwidth expansion over conventional designs. Experiments confirm broadband operation from 1480 to 1600 nm, achieving peak efficiency approaching 80\% and orbital angular momentum (OAM) purity up to 91.7\% under flat-top illumination, while suppressing edge effects and mitigating positional sensitivity and numerical-aperture (NA) dependence. As a proof of concept, we demonstrate direct conversion of zero-order Bessel beams into OAM Bessel (perfect vortex) beams with enhanced wavelength tunability, underscoring the versatility of this approach over diverse illumination conditions. This record-high performance establishes a practical and scalable pathway toward broadband, high-efficiency vortex generation, opening new opportunities across high-dimensional optical communications, advanced imaging, and quantum photonics.
\end{abstract}

\maketitle


\section{Introduction}
Vortex beams, defined by helical phase fronts and orbital angular momentum (OAM), are essential in many cutting-edge applications, including optical communications\cite{PhysRevLett.94.153901}, quantum information processing\cite{pnas.1616889113,WOS:000398262900021}, spin-orbit interactions of light\cite{WOS:001007507000001}, and super-resolution imaging\cite{Furhapter:05,PhysRevLett.97.163903}. They allow researchers and scientists to encode information in OAM\cite{Zhao-2024}, greatly expanding the capacity for high-dimensional data transmission and quantum entanglement\cite{RN501}.

Conventional approaches of generating vortex beam leverage phase modulation in real space\cite{RN501,RN506_nat_tunability,science.1210713,PhysRevLett.96.163905,Kotlyar:05,lpor.201600101,RN748,adma.201504532,WOS:000330182700019,Zhou2024OEA}, including dynamic phase and geometric phase\cite{PhysRevLett.96.163905}. 
Typical examples include spiral phase plate\cite{Kotlyar:05} and Q-plates\cite{Berry01111987,PhysRevLett.96.163905}. For the former case, it achieves spiral phase accumulation by varying the thickness or refractive index of the materials along the radial direction, and usually operates in the low-frequency range. For Q-plates operating in the visible wavelength range, a cross-polarized circularly polarized transmitted light gains the geometric Pancharatnam–Berry (PB) phase with a gradient along the axial direction. 
Although straightforward, these methods are bulky and highly sensitive to wavelength, limiting their versatility in many applications. 
Recently, a more compact platform, the so-called metasurfaces, has been developed to generate vortex beams in the visible and near-infrared wavelength range\cite{KG-2023,WOS:000330182700019, WOS:000515476500004}.
Metasurfaces, which rely on dynamic\cite{science.1210713,lpor.201600101,RN748} or geometric phases\cite{adma.201504532}, provide greater flexibility and wavelength independence, but 
are costly to fabricate and demand strict optical alignment to geometric centers and suffer from efficiency bottlenecks imposed by fabrication imperfections, especially on a short-wavelength scale. In addition, metasurfaces made of discrete meta-atoms can only modulate phase discontinuously\cite{RN748,science.1210713,Wang2024OEA}, limiting the quality of vortex beams.

More recently, momentum space polarization vortices centered at the bound states in the continuum (BICs) have been harnessed to generate vortex beams in a nonlocal metasurface\cite{RN269}.
This approach encodes vortex PB phase into cross-polarized light, opening a new avenue for light manipulation\cite{RN475,RN247,RN276,Chen2025,RN749}.
Except for BICs, degeneracy points (DPs) carrying polarization vortex have also been used to generate vortex beam in momentum space\cite{RN436,RN440}.
Unlike phase modulation in real space\cite{science.1210713,RN624}, momentum-space modulation eliminates the need for beam alignment with a geometric center and offers smoother phase modulation. However, the pursuit of a vortex platform concurrently delivering broadband operation, near-unity efficiency, and diffraction-limited purity—without positional sensitivity—has persisted as a critical challenge in photonics\cite{RN413}.
Specifically, the on-resonance conversion efficiency—defined at a specific wavevector—is limited by high-quality factor (Q-factor) resonances near the BICs. More importantly, the overall efficiency remains low, as the steep dispersion in the photonic crystal slabs (PCS) restricts the working bandwidth of the wavevector within the iso-frequency contour in momentum space\cite{RN815}. These restrictions are common in both BIC- and DP-based approaches, resulting in efficiency being highly dependent on the working wavelengths and wavevector, i.e. numerical apertures (NAs), which limits the tunability and flexibility of versatile optical platforms. To illustrate the limitations of existing designs, \autoref{tab:comparison_compact} summarizes representative nonlocal metasurface implementations, highlighting their narrow operational bandwidths, low overall efficiencies, and stringent NA constraints. This comparison motivates the need for a new platform capable of delivering high-efficiency, broadband operation without sacrificing beam quality or positional robustness.

In this study, we demonstrate broadband operational vortex beam generation with high efficiency through leveraging BICs and DPs synergistically in a nonlocal metasurface. We show that the dispersion, Q-factor, and topological states of polarization (SOPs) properties together in momentum space can be tailored through engineering the coupling between BICs and DPs. This approach effectively decouples efficiency limitations from the working wavelength and NAs seen in previous methods, enabling simultaneously high efficiency and wide tunability on wavelength. Numerical simulations show that the on-resonance efficiency and overall efficiency can reach as high as 96\% and 94\%, respectively, with OAM purity as high as 99.5\%. Besides, such a metasurface can generate vortex beam with the operation wavelength tunable from 1500 nm to 1600 nm. Following the theoretical design, we fabricated the relevant metasurfaces and successfully demonstrated the generation of vortex beam with overall efficiencies approaching 80\%, highlighting a significant improvement over previous studies. Furthermore, our measurements confirm that the overall efficiency remains reasonably high across the 1480 nm–1600 nm under various NAs while simultaneously showing an exceptional robustness to edge effect in periodic structures. This enables wavelength-tunable vortex generation with exceptional flexibility. Finally, by directly illuminating the structure with a zero-order Bessel beam, we experimentally generated efficient second-order Bessel beams (also known as perfect vortex beam\cite{RN544}), showcasing the versatility and flexibility of our design across diverse optical platforms.

\begin{table}[htbp]
    \centering
    \caption{Comparison of our design with previous non-local metasurface}
    \label{tab:comparison_compact}
    \begin{tabular}{|l|c|c|c|c|c|c|}
        \hline
        \textbf{\makecell{Ref.}} & \textbf{Design} & \textbf{\makecell{Wavelength \\ Band}} & \textbf{\makecell{Sim. \\ on-res \\ Eff.}} & \textbf{\makecell{Exp. \\ overall \\ Eff.}} & \textbf{\makecell{Operational \\ Bandwidth}} & \textbf{\makecell{Operational \\ NA}} \\
        \hline
        \makecell{\cite{RN269}} & PCS & Visible & - & - & narrow band & limited \\
        \hline
        \makecell{\cite{RN400}} & \makecell{Metal-backed \\ PCS} & Visible & 86\% & - & narrow band & limited \\
        \hline
        \makecell{\cite{RN436}} & \makecell{Metal-backed \\ PCS} & Visible & $<80\%$ & 13\% & narrow band & \makecell{NA=0.07 \\ (limited)} \\
        \hline
        \makecell{\cite{RN815}} & Metallic PCS & mm-wave & 41\% & 19.6\% & narrow band & \makecell{NA=0.4 \\ (limited)} \\
        \hline
        \makecell{\textcolor{red}{This} \\ \textcolor{red}{work}} & \makecell{Metal-backed \\ BZF PCS \\ + FP cavity} & Near-Infrared & \textcolor{red}{94\%} & \textcolor{red}{77\%} & \textcolor{red}{broadband} & \makecell{NA=0.1--0.42 \\ \textcolor{red}{(flexible)}} \\
        \hline
    \end{tabular}
\end{table}

\section{Results}
\subsection{Principle}

\begin{figure}
  \centering
  \includegraphics[width=0.8\textwidth]{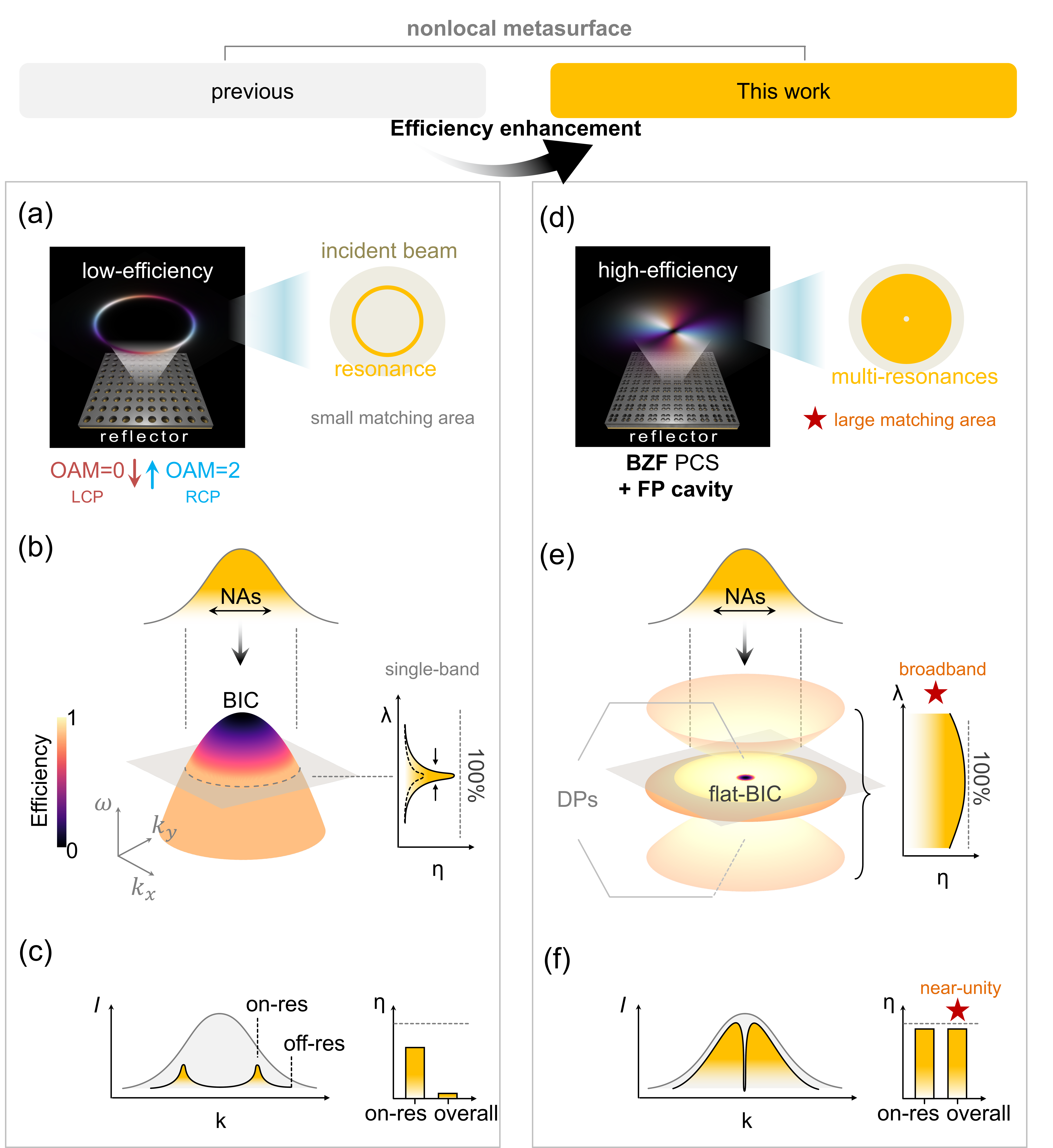} 
  \caption{
    \textbf{Overview of our theoretical principle.} (a–c) For conventional nonlocal metasurfaces based on a single BIC approach, the presence of a narrow efficiency ring in $(k_x,k_y)$ space (a) and steep modal dispersion (b) severely limits overall vortex beam conversion efficiency (c). (d–f) In contrast, our reflection-type BZF nonlocal metasurface (d) employs tailored band engineering to construct quasi-flat dispersion harnessing both BIC and DP modes (e), leading to broadband high-efficiency operation (f).
  }
  \label{fig-abstract}
\end{figure}

Vortex beams can be generated in the momentum space of nonlocal metasurfaces by exploiting the polarization vortices carried by guided resonances (GRs) in photonic crystal slabs (PCS). Such momentum-space modulation provides smooth, symmetry-protected phase profiles~\cite{RN748}, high tolerance to fabrication imperfections~\cite{adma.201504532}, without requiring precise beam alignment to a geometric center.

However, existing nonlocal schemes—typically based on a single BIC or a single DP—suffer from intrinsic efficiency bottlenecks. As illustrated in \autoref{fig-abstract}(a–c), steep modal dispersion confines momentum matching to a narrow efficiency ring in $(k_x,k_y)$ space, corresponding to the iso-frequency contour of the resonant mode. Only wavevectors lying within this thin ring achieve efficient vortex conversion; most of the incident energy falls outside this contour and remains unconverted. For BIC-based designs, the radiative Q-factor $Q_{\mathrm{rad}}\to\infty$ at the $\Gamma$ point suppresses scattering, creating a central low-efficiency void. For DP-based designs, orthogonal polarization branches cancel at $\Gamma$ (SOP cancellation, discussed later) similarly drive the conversion efficiency to zero. In both cases, the overall efficiency is low and highly sensitive to wavelength and NA, limiting practical applicability.

To analyze these effects quantitatively, we model the GR modulation as an anisotropic wave plate in momentum space with a position-dependent optical axis~\cite{RN269}. In local coordinates aligned with the resonance at azimuthal angle $\phi(k)$, the Jones matrix reads
\begin{equation}
\mathbf{J}_{\rm local} = \begin{bmatrix}
s_{\text{res-x}} + s & 0 \\
0 & s_{\text{res-y}} + s
\end{bmatrix}
\end{equation}
where $s$ denotes non-resonant background transmission, while $s_{\text{res-x}}$ and $ s_{\text{res-y}}$ capture resonance-induced polarization-dependent transmission for two orthogonal linear \(x\)- and \(y\)-polarizations, respectively. This form inherently separates non-resonant ($s$) and resonant ($s_{\text{res-x/y}}$) contributions. For a single resonance with linear polarization, $s_{\text{res-y}}$ can be zero. However, for multiple resonances with different polarization, the diagonal terms are both nonzero. Here we ignore the anti-diagonal term. A rigorous analysis for dual-mode with arbitrary polarization case is developed in the Supplementary Materials Section 1 to give the full matrix elements. Critically, our model concurrently accounts for contributions from two differently polarized resonances, departing from prior studies that enforced $s_{\text{res-y}}=0$.

We define the term $\Delta s_{\text{res}} = s_{\text{res-x}}-s_{\text{res-y}}$ to characterize the polarization dependence of the resonance. Notably, this parameter plays a critical role in vortex beam generation by representing the intensity of the effect of PB phase. To show that, we perform coordinate rotation $\mathbf{R}(\phi)$ to global $(x,y)$ system followed by helical basis transformation. Therefore, for right-circular input $\ket{R}$, the output field components derive as\cite{RN269}:
\begin{align}
\ket{E_{\text{out}}} &= \frac{1}{2}(s_{\text{res-x}}+s_{\text{res-y}}+2s)\ket{R} + \frac{1}{2}\Delta s_{\text{res}}e^{-2i\phi}\ket{L}
\label{VBG-main_eq}
\end{align}
Here, the cross-polarized component $\frac{1}{2}\Delta s_{\text{res}} e^{-2i\phi}\ket{L}$ encodes the PB phase $\Phi_{\rm PB} = -2\phi$ via its $e^{-2i\phi}$ factor\cite{Berry01111987}. We can define the cross-polarization conversion efficiency as $\eta=|\bra{L} \mathbf{J}_{\rm circ} \ket{R}|^2=|\frac{1}{2} \Delta s_{\mathrm{res}}|^2$. Introduce the topological encoding $\phi(\theta_k) = q\times\theta_k$, where $\theta_k$ is the momentum-space azimuthal angle and $q$ is the topology polarization winding number of the GR modes\cite{RN247,RN276}. This produces a vortex beam with OAM $2\times q$, whose left‐circular component satisfies $\braket{L|E_{\text{out}}} \propto e^{-2iq\theta_k}$.
Importantly, in a \(C_4\)‑symmetric metasurface, the existence of polarization vortices in momentum space is enforced by symmetry alone\cite{RN276}. Such polarization topology emerges as a global property of the system. The key to vortex beam generation in $C_4$-symmetric systems does not rely on BICs\cite{RN269} or DPs\cite{RN317}, but rather on achieving cross-polarization conversion, which naturally leads to vortex phase—a distinction not previously clarified before (see Supplementary Materials Section 5 for numerical validation)

In general, high‑efficiency vortex‑beam generation is synonymous with low‑\(Q_\text{rad}\), strongly scattered resonances\cite{RN400}. For a single linear-polarized resonance, i.e. $s_{\text{res-x}}=s_{\text{res}},s_{\text{res-y}}=0$, the resonance strength $\Delta s_{\text{res}}=s_{\text{res}}$ under external excitation is governed by temporal coupled mode theory (TCMT)\cite{RN239}:  $s_{\text{res}} = -\frac{d_s^2+d_p^2}{\gamma_{\text{rad}}+\gamma_{\text{abs}}-i(\omega-\omega_0)}$,
where $\omega$ is the excitation frequency and $\omega_0$ is the resonance frequency. $\gamma_{\text{rad}}$ and $\gamma_{\text{abs}}$ describe the radiative and absorption loss of the resonance, respectively. The radiation coupling is described by $d_i$ (here we ignore the phase factor for complex coefficients $d_s$ and $d_p$). Notably, the intensity of the scattering field can always be strongly enhanced using a reflection-type structure; hence, we adopt this configuration for subsequent discussions. In a reflection‑type PCS, energy conservation dictates \(\lvert d_s\rvert^2 + \lvert d_p\rvert^2=2\gamma_{\rm rad}\)\cite{RN413,RN400}, so on resonance (\(\omega=\omega_0\)) the cross‑conversion efficiency simplifies to $\eta =|\frac{1}{2} s_{\mathrm{res}}|^2 \sim \left(Q_{\mathrm{abs}}/( Q_{\mathrm{rad}}+Q_{\mathrm{abs}})\right)^{2}$, where \(Q_{\rm rad}\) and \(Q_{\rm abs}\) are the radiative and absorptive Q-factors.  While \(Q_{\rm abs}\) is governed by material loss and thus difficult to reduce, lowering \(Q_{\rm rad}\) (i.e. engineering stronger radiative coupling) directly boosts \(\eta\).  This insight guides our design of low‑\(Q_\text{rad}\) resonances to achieve the high conversion efficiencies demonstrated above.

In \autoref{fig-abstract}(a–c), we show why neither BIC‑ nor DP‑based schemes can yield high‐efficiency vortex beams.  
For BIC-based approaches, the modes indeed carry robust polarization vortices ($s_{\text{res-x}} \neq s_{\text{res-y}}$), but their radiative Q-factor diverges near the \(\Gamma\) point (\(Q_{\rm rad}\to\infty\)). Hence the resonance amplitude $s_{\rm res}$ approaches $0$, and the cross‑conversion term \(\Delta s_{\rm res}\) vanishes, producing a pronounced large conversion “black hole” near the \(\Gamma\) point, as depicted in \autoref{fig-abstract}(b).
On the other hand, DPs avoid the divergence of \(Q_{\text{rad}}\) that is typically associated with polarization singularities by degeneracy-induced polarization independence at the \(\Gamma\) point. However, such a degeneracy also leads to suppressed cross-polarization conversion, resulting in a large low-efficiency area near the \(\Gamma\) point as well. For instance, consider DPs with $+1$ topological charge, their split bands, each carrying polarization vortices with opposite SOPs, can only be azimuthal polarized (AP) or radial polarized (RP) under $C_4$ symmetries, as later discussed in \autoref{fig-principle}(b-d). At the \(\Gamma\) point, suppressed cross-conversion occurs due to the competition between these two completely orthogonal-polarized SOPs, resulting in polarization-independent modes with $s_{\text{res-x}}=s_{\text{res-y}}, \to \Delta s_{\text{res}} =0$, and again quenching cross‑polarization conversion (cf. \autoref{VBG-main_eq}). We term this mechanism as SOPs cancellation. A detailed discussion using the dual-resonance TCMT is provided in the Supplementary Materials Section 1.

Moreover, besides the resonance problem, both BIC- and DP-related GRs share another common issue: dispersion\cite{RN413, RN400,RN631}.
Although the theoretical on-resonance efficiency can reach 100\% in reflection-type structures, the overall efficiency is still limited by mode dispersion. Dispersion induces a momentum mismatch between the excitation beams and the resonant modes. Consequently, the iso-frequency efficiency contours assume ring-like shapes with a null center for conversion. As shown in the right panel of \autoref{fig-abstract}(b), this momentum-space mismatch implies that the conversion region is highly sensitive to the incident wavelength and can operate effectively only for specific wavevectors (i.e., NAs) and wavelengths that coincide with the momentum-space ring. Consequently, even when the on‑resonance efficiency reaches its theoretical maximum, the overall efficiency remains relatively low across a broadband wavelength range, as illustrated in \autoref{fig-abstract}(c). More detailed quantitative analysis is provided in Supplementary Materials Section 2.

Therefore, as shown in previous studies\cite{RN430,RN631,RN400,RN468,RN413,RN269,RN629}, the conversion efficiency displays a significant narrow efficiency ring (\autoref{fig-abstract}(a)), caused by the aforementioned two synergistic effects. This renders the conventional approaches not suitable for achieving  highly efficient and broadband vortex beam generation. 

To overcome these intrinsic barriers, we introduce an on‐demand band‐engineering strategy that simultaneously tailors dispersion, radiative $Q$‑factor, and polarization topology of GRs. Our work (\autoref{fig-abstract}(d-f)) demonstrates an on-demand band engineering for both BIC- and DP-related modes. As shown in \autoref{fig-abstract}(e), by engineering GRs with quasi-flat dispersion and exceptionally strong scattering radiation (low Q-factor), the momentum match between the incident beam and the resonance mode is satisfied(\autoref{fig-abstract}(d)), which leads to high overall efficiency for broadband vortex beam generation, as shown in \autoref{fig-abstract}(f). Specifically, we strategically introduce mode coupling to establish a link between BICs and multiple DPs, as illustrated in \autoref{fig-principle}.

\subsection{Numerical Implementation}
\begin{figure}
  \centering
  \includegraphics[width=1.0\textwidth]{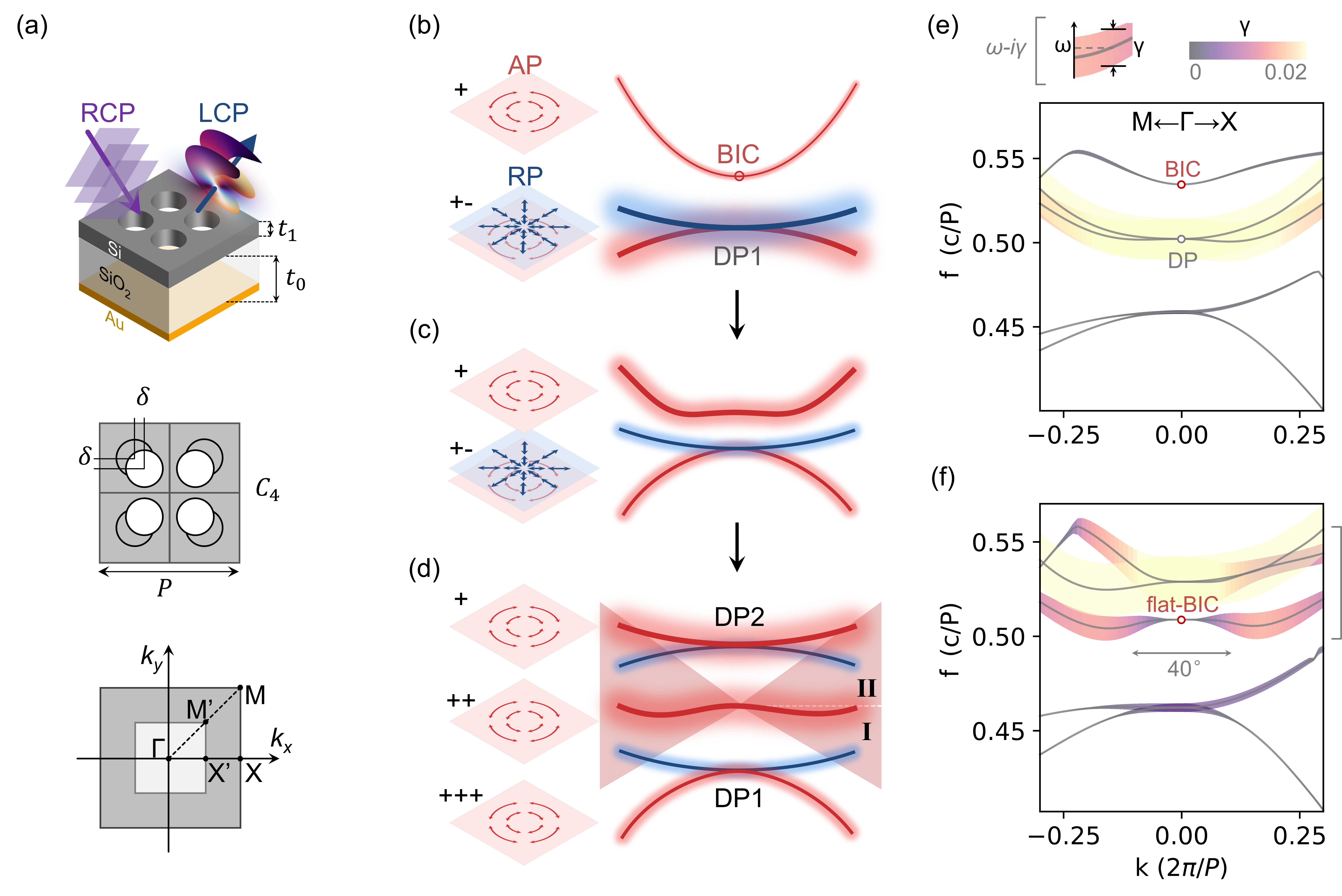} 
  \caption{
    \textbf{Overview of our implementation.} (a) Top panel: BZF PCS structure converting an incident right-handed circularly polarized plane wave into a left-handed circularly polarized vortex beam. Middle panel: Periodic perturbation maintaining \(C_4\) symmetry. Bottom panel Brillouin zone folding in momentum space. (b–d) Band-structure engineering diagrams; glowing regions represent mode bandwidth (\(Q\) factor), red/blue colors denote AP/RP SOPs respectively. Broadband high-efficiency regimes I and II are formed in (d) through synergistic effects of BIC- and DP-related modes. (e,f) Simulated band structures for conventional and optimal designs, with line thickness and color indicating mode bandwidth.
  }
  \label{fig-principle}
\end{figure}

Our implementation, illustrated in \autoref{fig-principle}(a), employs a reflection-type PCS vortex generator: an air-hole PCS in silicon atop a \ce{SiO_2} spacer backed by a gold mirror. The high-index PCS confines the GR modes, while the \ce{SiO2} spacer decouples the fields from the Au substrate, significantly reducing Ohmic losses. Furthermore, this configuration forms a FP (Fabry-Perot) cavity between the top PCS and the gold substrate (\autoref{fig-principle}(b)), where the spacer thickness $t_0$ allows independent tuning of the FP resonance wavelength while suppressing Au absorption.

Periodic perturbations are introduced into the PCS lattice, preserving \(C_4\) symmetry (\autoref{fig-principle}(a), middle panel), to enable Brillouin zone folding (BZF)~\cite{RN219,RN719,RN245,RN150}, generating GR modes with controllable wavelength and $Q$ factor (\autoref{fig-principle}(a), bottom panel). In this hybrid system, the BIC arises from a high-\(Q\) GR band of the PCS, while DPs are contributed by FP cavity modes with two low-\(Q\) bands. All modes carry a topological charge of \(+1\) but exhibit different SOPs (AP vs. RP).

To systematically achieve efficient vortex generation with multidimensional tunability, we start with a simplified system including DP\textsubscript{1} and BIC singularities. The energy bands that relate with DP\textsubscript{1}, BIC, DP\textsubscript{2} are labeled from bottom to top as \(\text{FP}_1^{\text{A}}\), \(\text{FP}_1^{\text{R}}\), \(\text{GR}^{\text{A}}\), \(\text{FP}_2^{\text{R}}\), and \(\text{FP}_2^{\text{A}}\), where the superscripts A/R denote the azimuthal and radial SOPs, respectively (see left panel in \autoref{fig-principle}(b)).

Initially, as illustrated in \autoref{fig-principle}(b), neither BIC nor DP efficiently generates vortex beams—the former is due to excessive Q-factor restricting bandwidth, and the latter is due to SOPs cancellation between $\text{FP}_1^{\text{A}}$ and $\text{FP}_1^{\text{R}}$ bands (left panel).
Furthermore, the steep dispersion of the initial structure permits only a narrow tolerance in wavelength and wavevector.

Subsequently, we tune the wavelength of the first DP (DP\(_1\)) to facilitate the interaction between \(\text{FP}_1^{\text{A}}\) and \(\text{GR}^{\text{A}}\) (dashed line in \autoref{fig-principle}(c)). This tuning establishes strong coupling between \(\text{GR}^{\text{A}}\) and \(\text{FP}_1^{\text{A}}\)\cite{RN150}, resulting in an avoided crossing that flattens the dispersion of \(\text{GR}^{\text{A}}\) around the BIC. Simultaneously, the off-\(\Gamma\) \(Q\)-factor is reduced to FP levels, dramatically enhancing the conversion efficiency while maintaining the strongly polarization-dependent response. Notably, the low \(Q\)-factor signifies a short resonance lifetime, which enhances the system's robustness against edge effects in periodic structures, as the resonances cannot travel far before radiating. This behavior is later confirmed by our experimental results.
The engineered flat dispersion (\autoref{fig-principle}(c)) broadens the NA tuning range and establishes a flat, triangular high-efficiency region (indicated by the red background) around the BIC frequency. However, this single mode approach can only generate vortex beams within a narrowband wavelength range.

To further expand the efficiency regime, we introduce a secondary DP (DP\(_2\)), as shown in \autoref{fig-principle}(d).
First, we adjust the wavelength of the \(\text{GR}^{\text{A}}\)-\(\text{FP}_1^{\text{A}}\) mode to mitigate the disruptive effects of \(\text{FP}_1^{\text{R}}\) in the lower-frequency Region I. Then, we promote an interaction between \(\text{GR}^{\text{A}}\) and \(\text{FP}_2^{\text{A}}\). This interaction suppresses the \(Q\)-factor of \(\text{FP}_2^{\text{A}}\) and benefits radiation in the azimuthal SOP over the radial SOP (\(\text{FP}_2^{\text{R}}\)). Consequently, a higher-frequency Region II is created that supports efficient cross-polarized conversion. The avoided crossing between \(\text{GR}^{\text{A}}\) and \(\text{FP}_2^{\text{A}}\) further flattens the BIC band at larger wavevectors.
As shown in \autoref{fig-principle}(d), such a bidirectional band engineering, driven by the synergistic contributions of $\text{GR}^{\text{A}}$, $\text{FP}_2^{\text{A}}$ and $\text{FP}_1^{\text{A}}$, dominates the radiation in azimuthal SOPs (left panel) and establishes a special hourglass-shaped high-efficiency regime I/II, and provides unprecedented wavelength/NA tunability for vortex beam generation.

Optimized parameters through full-wave simulations yield: \ce{Si} thickness $t_1=157$ nm, \ce{SiO_2} thickness $t_0=545$ nm, hole radius $r=133$ nm, hole offset $\delta=52$ nm, and lattice constant $P=2a=1570$ nm. The system behavior under various geometric parameters is discussed in the Supplementary Materials Section 7.
Note that material absorption plays a critical role in conversion efficiency, so we assign an imaginary refractive‐index component of \(10^{-4}\)—a realistic value for chemical vapor deposition (CVD) ‐deposited films.  Specifically, we set \(n_{\rm Si} = 3.58 + 10^{-4}i\) and \(n_{\rm SiO_2} = 1.45 + 10^{-4}i\).
For comparison, a reference design—representative of the typical efficiency limitations reported in previous studies—is obtained by setting $\delta' = 0.7\delta$. The band structures of both the optimal and alternative designs are shown in \autoref{fig-principle}(e, f), respectively, successfully validating our proposed band-engineering concept. Further details are provided in Section 3 of the Supplementary Materials.

The strong coupling between BIC and DP modes is evidenced by an avoided crossing in the parameter space (Supplementary Materials Section 7), resulting in a system that exhibits dominant AP contributions from \(\text{GR}^{\text{A}}\) and \(\text{FP}_2^{\text{A}}\). This enables broadband vortex generation, with the BIC-based \(\text{GR}^{\text{A}}\) governing efficient conversion below 1550 nm and the DP-based \(\text{FP}_2^{\text{A}}\) dominating above 1550 nm, as illustrated by the hourglass-shaped high-efficiency region in \autoref{fig-sim_rsl}(d). Detailed SOP characteristics for each band are provided in the Supplementary Materials Section 6. Building on the above mode‐coupling analysis, we now examine the performance contrast between the optimized and reference structures.

\begin{figure}
  \centering
  \includegraphics[width=1.0\textwidth]{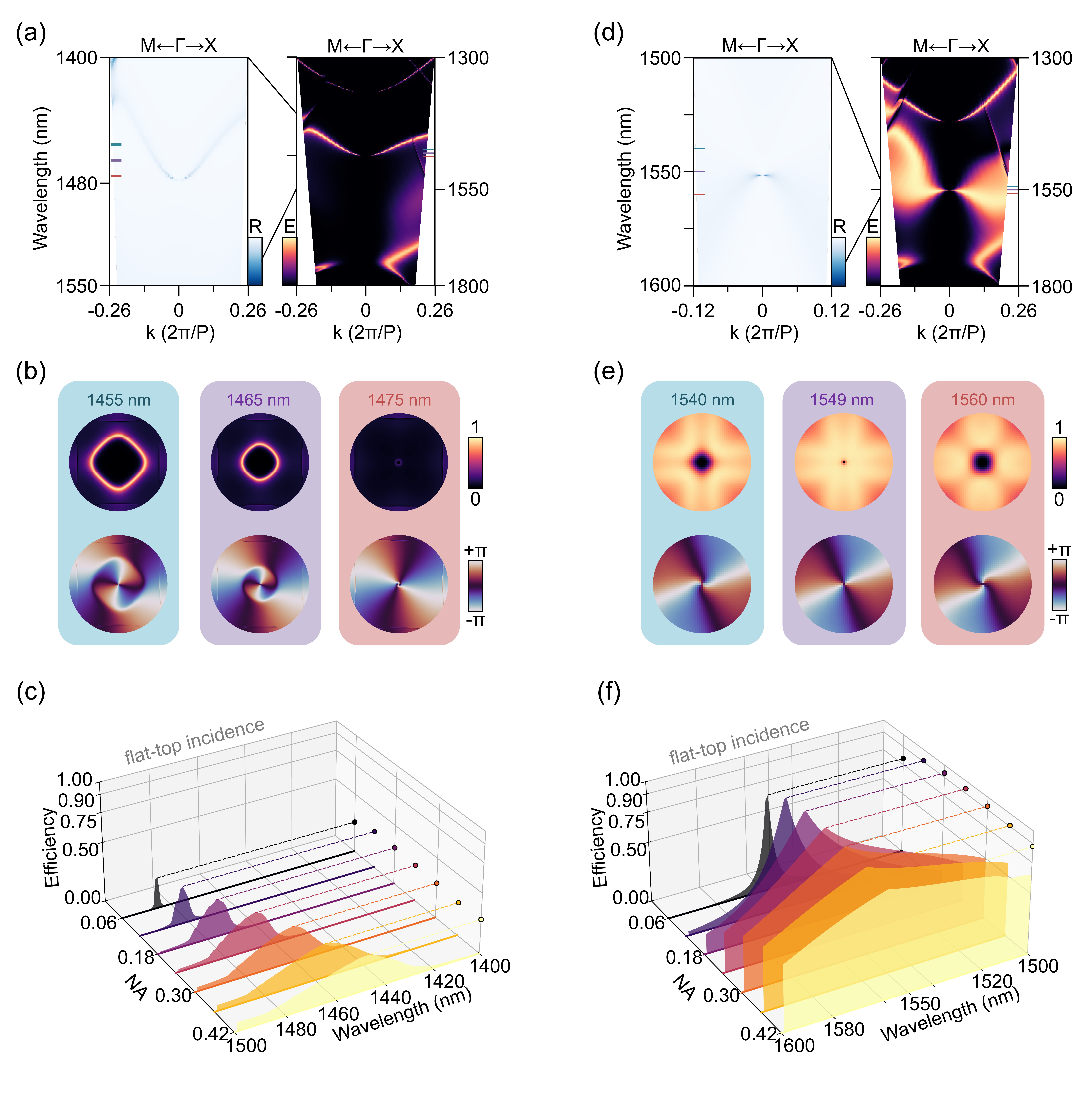} 
  \caption{
    \textbf{Numerical results of our design.} (a) Detailed spectrum of conversion efficiency \(\eta(\omega,\mathbf{k})\) for the comparson structure. Left: s-polarized reflection spectrum (\(R_s\)); right: cross-polarized conversion efficiency spectrum (\(\eta\)). (b) Colored panels: cross-sectional efficiency and phase at different wavelengths under NA=0.42. (c) Simulated overall efficiency of flat-top-beam illumination under different NA. (d–f) Same as (a–c) but for optimal structure.
  }
  \label{fig-sim_rsl}
\end{figure}

For the reference design (\autoref{fig-sim_rsl}(a–c)), a BIC appears near 1480~nm and a DP near 1500~nm, as shown in \autoref{fig-principle}(e). The efficiency $\eta$ shows only limited peaks around the BIC, consistent with the resonance linewidths of $R_s$. At 1500~nm, DP conversion is suppressed due to SOP cancellation. As shown in \autoref{fig-sim_rsl}(b), a narrow efficiency ring emerges in momentum space, in agreement with dispersion-limited matching calculated in Supplementary Section 2.

In contrast, the optimized design (\autoref{fig-sim_rsl}(d–f)) achieves significantly enhanced broadband efficiency. The \(+1\) topological-charge BIC, relocated near 1550~nm, enables maximal radiative resonance utilization across 1550–1600~nm, reaching up to \(\sim96\%\) on-resonance efficiency at (1549~nm, 3°). Meanwhile, the DP at 1500~nm enhances conversion in the 1500–1550~nm range. Together, these coupled modes form an hourglass-shaped high-efficiency region from 1500 to 1600~nm (\autoref{fig-sim_rsl}(d)), outperforming the conventional structure in \autoref{fig-sim_rsl}(a). 
Importantly, the preserved $C_4$ symmetry ensures azimuthally uniform conversion profiles near center, while the widely flattened dispersion minimizes the influence of central efficiency voids (seen in iso-contours in colored panels). Compare between \autoref{fig-sim_rsl}(b, e), since the narrow conversion ring has been greatly expanded, the cross-shaped conversion efficiency near the edges can be effectively eliminated in practical applications by limiting the beam's focus angle (i.e. NA), to achieve optimal azimuthal uniformity and OAM purity without sacrificing efficiency.
\autoref{fig-sim_rsl}(c, f) show the simulated overall efficiency under flat-top beam incidence, obtained by averaging the maps in \autoref{fig-sim_rsl}(b, e). Compared to conventional nonlocal metasurfaces (\autoref{fig-sim_rsl}(c)), our design achieves more than threefold higher efficiency with flat-top beams (\autoref{fig-sim_rsl}(f)), fourfold higher efficiency with Gaussian beams, over fifteenfold broader bandwidth with Bessel beams, and far greater tolerance to varying NA. The overall efficiency reaches 94\% at 1549 nm within a NA=0.21, with near-unity (up to 99.5\%) OAM purity and smooth phase manipulation. Full comparisons for different beam types are provided in Supplementary Section 4, while OAM quality is evaluated via LG-mode decomposition in Section 9.

Overall, our design surpasses conventional nonlocal metasurfaces in both efficiency and tunability, offering robust momentum-space generation and strong resilience against edge effects, as further confirmed experimentally.

\subsection{Experimental Demonstration}

\begin{figure}
  \centering
  \includegraphics[width=1.0\textwidth]{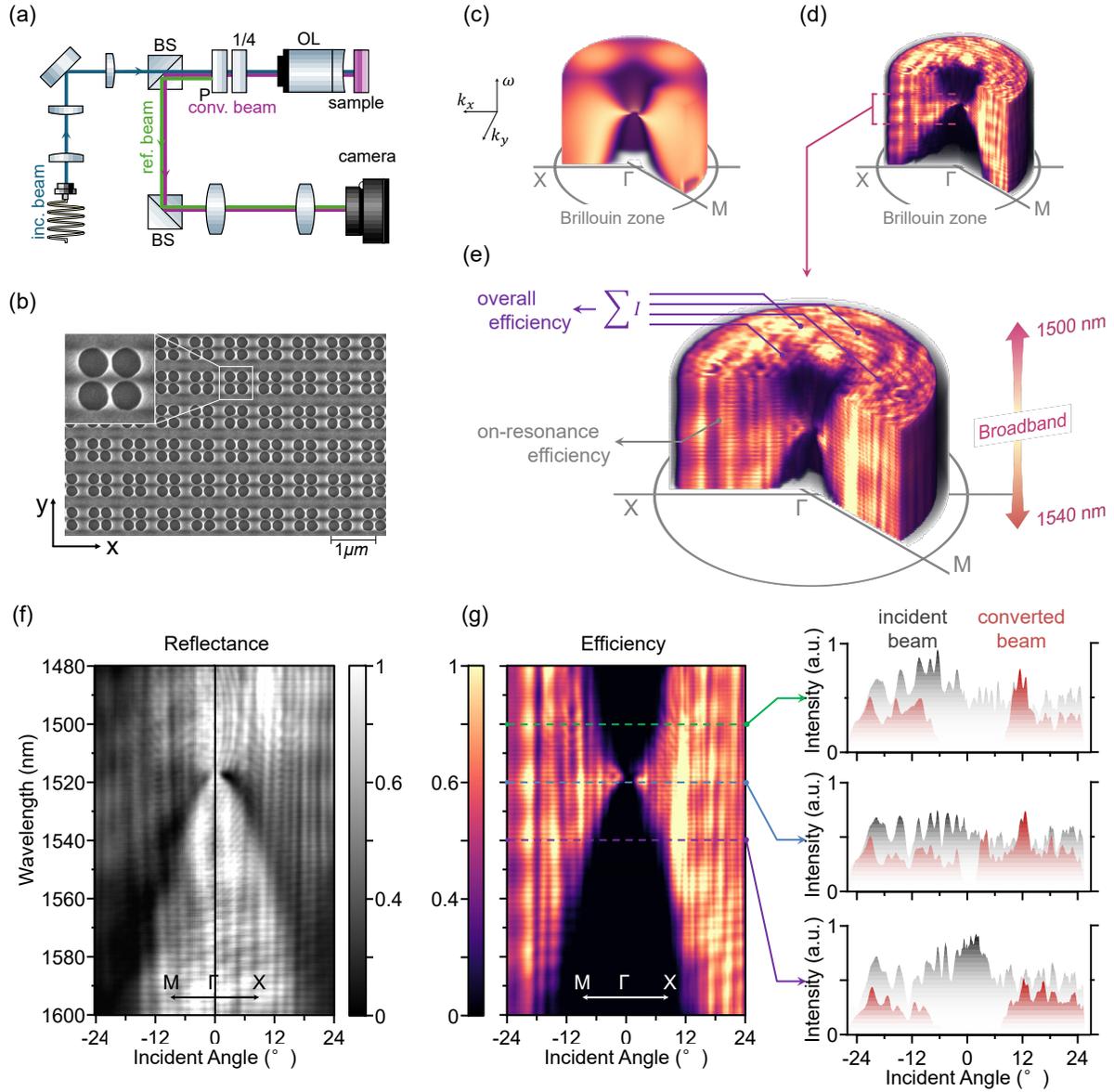} 
  \caption{
  (a) Schematic of the momentum-space imaging optical setup employed in the experiment. (b) Top‐view SEM micrograph of the fabricated metasurface. (c-e) Three‐dimensional full $\omega$–$\mathbf{k}$ dispersion of the cross‐polarization conversion efficiency: (c) simulated results; (d-e) experimental measurements, with (e) highlighting broadband and high overall conversion efficiency across the 1500–1540 nm range. (f) Measured s-polarized reflection spectrum $R_s$. (g) Cross‐polarized conversion efficiency spectrum. Colored dashed lines in (g) mark the wavelengths—1500 nm (green), 1520 nm (blue) and 1540 nm (purple)—for which the right subpanels show the absolute intensities of the incident (black) and converted (red) beams.
  }
  \label{fig-exp_measurements}
\end{figure}

To confirm the superior performance predicted by our simulations, we experimentally demonstrate the vortex beam generation using our optimized structural parameters.
A \ce{Si}/\ce{SiO_2} multilayer was grown through plasma-enhanced chemical vapor deposition (PECVD). PCS structures with four air holes in a super unit cell are fabricated with electron beam lithography followed by inductively coupled plasma reactive ion etching, as shown in \autoref{fig-exp_measurements}(b). More details of the fabrication process can be found in Methods.

A home-built reflection-type Fourier-optics imaging system with dual operation modes (imaging and interferometry) was developed for optical characterization, as schematically illustrated in \autoref{fig-exp_measurements}(a). The system is composed of three functional modules: (i) beam generation, (ii) vortex beam conversion, and (iii) converted beam collection/imaging. In imaging mode, left-handed circularly polarized light from a laser source (generated by a linear polarizer and quarter-wave plate (JCOPTIX, China)) was focused onto the PCS through an objective lens with NA=0.42, which is consistent with our simulation NA range. The cross-polarized reflected light underwent Fourier transformation through the same optical path before being captured by an InGaAs CMOS camera (JCOPTIX, China) in the beam collection/imaging module.

For phase characterization, a self-referenced interferometric configuration was implemented. The front-surface reflection from polarizer P served as the reference beam, which interfered with the generated vortex beam in the collection path. By adjusting the yaw angle of polarizer P, three interference modes were achieved: reference-free (for conversion efficiency measurement), tilted-reference (for oblique phase imaging), and collinear-reference (for coaxial phase retrieval). Wavelength-dependent spectral features were obtained by scanning the laser wavelength, revealing the full \(\omega\)-\(\mathbf{k}\) dispersion relation (\autoref{fig-exp_measurements}(c-e), also see supplementary video1 for details). The broadband high-efficiency, and the difference between overall efficiency and the on-resonance efficiency is demonstrated in \autoref{fig-exp_measurements}(e). The s-polarized reflection spectra (obtained by rotating the quarter-wave plate) and cross-polarized conversion efficiency profiles are shown in \autoref{fig-exp_measurements}(f, g), respectively, demonstrating a broadband "hourglass" region of high efficiency consistent with theoretical predictions.
The BIC is located around 1520 nm, slightly deviating from the numerical simulations primarily due to discrepancies between the actual refractive index of the structure and the refractive index used in the simulations. This mismatch leads to a globally slight blueshift in the spectrum, but does not significantly affect the structural performance (see Supplementary Materials Section 7 for details).
The right subpanels of \autoref{fig-exp_measurements}(g) show the absolute intensities of the incident non-vortex beam (black) and the converted vortex beam (red) around 1500, 1520, and 1540 nm. In particular, the mid-subpanel of 1520 nm demonstrates remarkable robustness of the intensity converted beam against variations in incident angle, which underpins the device’s high overall efficiency. Note that, since we sample only along two one-dimensional cuts—along the $\Gamma$–X and $\Gamma$-M directions—through the imaged two-dimensional momentum-space intensity distribution, diffraction-speckle fluctuations are observed.

\begin{figure}
  \centering
  \includegraphics[width=1.0\textwidth]{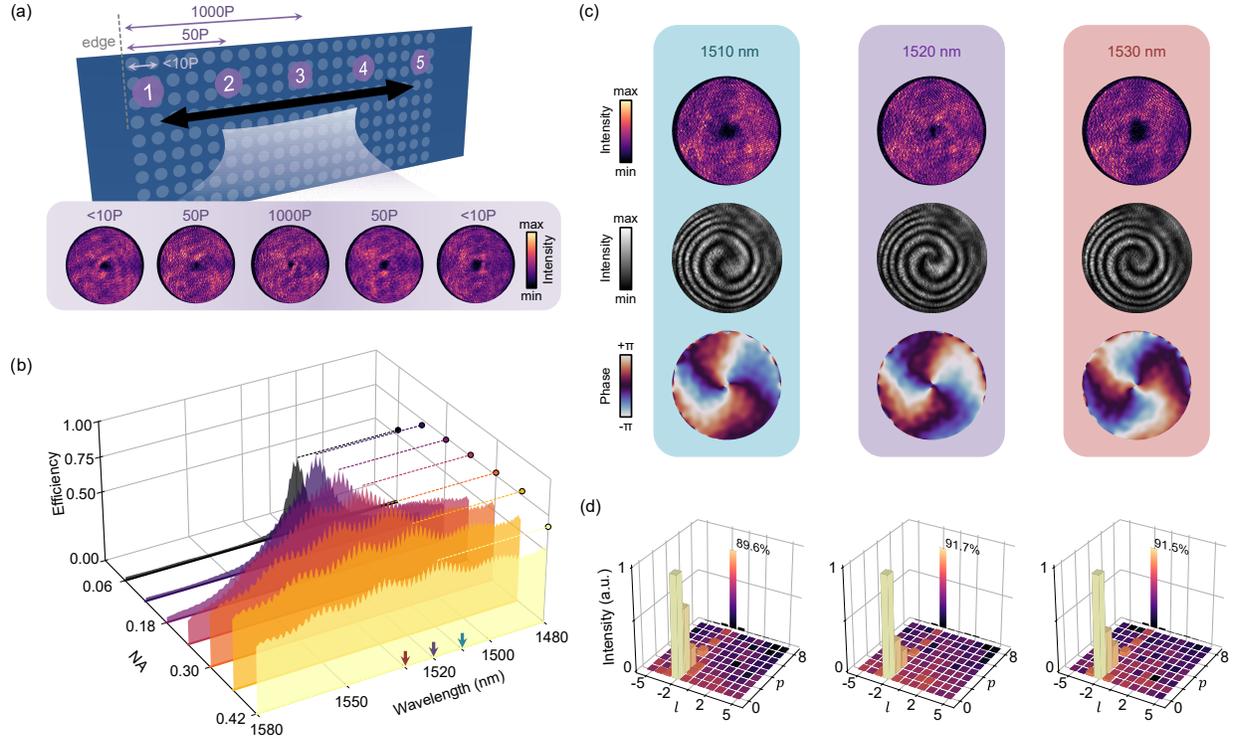} 
  \caption{
  (a) Measured cross-polarization converted beam profile at different illumination positions. (b) Measured overall averaged efficiency under different wavelength and NA. (c) Measured generated beam profile. From top to bottom: intensity, interference pattern and extracted phase.  (d) LG mode decomposition of the converted vortex beam under NA=0.21 at 1510, 1520, and 1530 nm. Inset: OAM ($l$) distribution, where the purity of $l=-2$ can reach as high as 91.7\% at 1520 nm.
  }
  \label{fig-exp_rsl}
\end{figure}

Next, we confirm the robustness of our design by verifying the decentralization of momentum-space vortex beam generation and examining its dependence on wavelength and NAs.
The decentralization of vortex generation was first verified by probing regions from the structure’s edge (\(<10P\)) to its center (\autoref{fig-exp_rsl}(a)). Considered the \(3P\)-sized beam spot in NA=0.42, consistent vortex beam generation across all positions confirmed the position-independent operation and robustness against edge effects. As we have predicted, the disappearance of edge effects in the PCS is due to the low $Q$-factor of the emission mode we utilize, causing the resonance to radiate spontaneously before it can propagate to the edge.

The wavelength- and NA-dependent efficiency was then systematically investigated (\autoref{fig-exp_rsl}(b)). By illuminating the sample with momentum-space flat-top beam, the overall efficiency was calculated by integrating the total beam intensity from the acquired images and comparing it to a reference beam reflected by an Au mirror. More details are presented in Supplementary Materials Section 4. The efficiency shows an exceptional robustness against wavelengths under NA=0.42, and the efficiency and quality can be further enhanced by reducing the NA. At NA=0.24, the structure shows improved overall conversion efficiency over $60\%$ with a ±25 nm bandwidth, peaking at 77\% at the central wavelength of 1520 nm, as shown in \autoref{fig-exp_rsl}(b). 
The reduction in experimental efficiency compared to numerical predictions can be attributed to three main factors. First, the roughness of the fabricated sample induces scattering losses, reducing the reflectance. Second, fabrication errors cause deviations from the optimal geometry. Third, interference speckle in the imaged beam introduces errors in the intensity integration. By focusing the beam to minimize the influence of interference speckles, efficiencies exceeding 80\% under NA=0.42 were achieved. Further details on this efficiency measurement are provided in the Supplementary Materials Section 4.
Overall, the efficiency was preserved across a wide range of NAs and wavelengths, demonstrating the tunability and flexibility of our platform for diverse optical configurations.

It is noteworthy that the light intensity distribution of beams commonly used in imaging and computational applications is uniform in momentum space (i.e., momentum-space flat-top beams). Consequently, the calculated efficiency corresponds to an unweighted average over momentum space, as illustrated in \autoref{fig-sim_rsl}(e, f). In contrast, another widely employed beam type—Gaussian beams—exhibits a generation efficiency that deviates from that of flat-top beams by a Gaussian weighting factor\cite{RN436}. Given our structure’s robust performance across various wavevectors, the overall efficiency and quality remain comparable between these beam types. One can find more details in Supplementary Materials Section 4.

Finally, the vortex phase and the quality of the generated beam in momentum-space were analyzed through interferometric imaging (\autoref{fig-exp_rsl}(c)). The reference beam for interference, generated by front-surface reflections from polarizer P, was combined with the converted vortex beam for phase analysis via Fourier methods\cite{RN673} (see Supplementary Materials Section 9 for details). Three critical aspects were demonstrated: Firstly, the robustness of high-quality vortex generation across wavelengths was reconfirmed, with measured intensity profiles (first row) showing excellent agreement with simulated conversion efficiencies. Secondly, collinear interference with a reference plane wave (second row) exhibited two distinct helical arms in the fringe pattern, unambiguously verifying the topological charge of OAM = 2. Thirdly, reconstructed phase distributions (third row) revealed smooth \(0-4\pi\) azimuthal variations, confirming high-fidelity vortex phase generation.
To quantify beam quality, Laguerre-Gaussian (LG) mode decomposition was performed\cite{RN675,RN499}. The decomposition component is illustrated in \autoref{fig-exp_rsl}(d), showing $>90\%$ OAM purity under NA=0.21 (see Supplementary Materials Section 9 for details). Here, the OAM purity did not reach theoretical prediction ($>$99\%) due to experimental measurement errors and the non-azimuthally uniformity near the edge of high NA components. Notably, reducing the NA can improve both beam quality and efficiency, as detailed about LG mode decomposition provided in the Supplementary Materials Section 9.
These results collectively validate the high quality and adaptability of our vortex generation mechanism.

\begin{figure}
  \centering
  \includegraphics[width=1.0\textwidth]{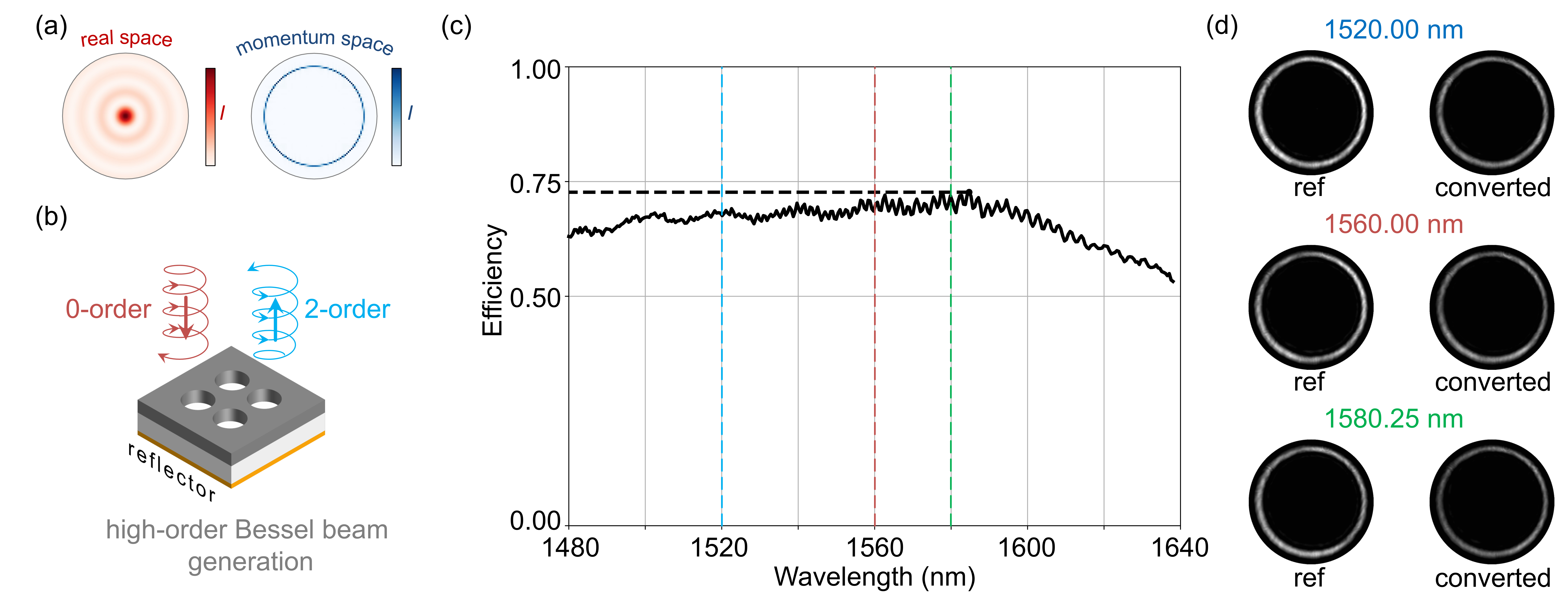} 
  \caption{
  (a) Beam profile in real space (left) and momentum space (right) for zero-order Bessel beam. (b) Generation of high-order Bessel beam by direct illuminate the circular polarized zero-order Bessel beam at NA=0.42. (c) Measured conversion efficiency at various wavelengths. (d) Beam profiles at three different wavelengths, exhibiting high-quality, azimuthally uniform annular distribution.
  }
  \label{fig-bessel}
\end{figure}
In addition to vortex beams generated under flat-top illumination, we investigated the performance of Bessel-beam incidence due to their distinctive momentum-space characteristics\cite{RN501}. Although Bessel beams exhibit complex intensity profiles in real space, their momentum-space representations reduce to simple annular distributions, as shown in \autoref{fig-bessel}(a). This property makes them particularly suitable for momentum-space light-field modulation.

As illustrated in \autoref{fig-bessel}(b), zero-order Bessel beams were generated using axicon lenses, and subsequently converted into high-quality higher-order Bessel vortex beams by illuminating the same sample at NA=0.42. The reflected light, analyzed with the circular polarizer in \autoref{fig-exp_measurements}(a), revealed a second-order Bessel vortex beam carrying OAM of \( \ell = +2 \). Conversion efficiency was quantified by integrating the unfocused direct image of the beams. The annular profile in momentum space enabled selective utilization of the regions with the highest efficiency, thereby improving performance metrics such as generation efficiency, wavelength tunability, and NA tolerance. 

As shown in \autoref{fig-bessel}(c), the efficiency remains highly robust across wavelengths (1480 nm - 1600 nm), even under high NA conditions (NA = 0.42, where optimal NA is around 0.2 by simulation), in good agreement with theoretical predictions. Moreover, the annular intensity profile observed in \autoref{fig-bessel}(d)—characteristic of a "perfect vortex beam"—confirms the structure’s ability to reliably impart vortex phase across diverse beam configurations without altering energy distributions. Detailed experimental setups are provided in Supplementary Section 8.

\section{Conclusion}
In summary, we propose an innovative approach to achieve broadband operational vortex beam generation with high efficiency and high purity with a nonlocal metasurface. By establishing a hybrid coupling mechanism that links BIC with two DPs, we show that such a metasurface allows for precise, on-demand control over dispersion, Q-factor, and polarization state of the guided resonances. Simulation results show that such vortex beam generation can operate across a broad wavelength range (1500 nm -- 1600 nm) while preserving near unity conversion efficiency. We demonstrated such vortex beam generation experimentally by fabricating and characterizing the designed metasurfaces. Measurement results show the overall efficiency can approach 80\% under both Gaussian and Bessel beams’ illumination. We also confirmed enhanced generation quality and tunability across the 1480–1600 nm range under varied NA. Besides, we experimentally observed OAM purities of up to 91.7\% and exceptional robust performance against edge effects in finite-size arrays. This approach surpasses the intrinsic trade-offs of previous techniques, establishing a practical pathway for broadband, high-efficiency momentum-space vortex beam generation in real-world applications.

\section{Acknowledgments}
This work was supported by the Natural Science Foundation of Sichuan Province (Grant No. 2024NSFSC0460) and the National Natural Science Foundation of China (Grant No. 12274269, 12474377)

\section{Methods}

\subsection*{Theoretical analysis}
Please see the Supplementary Materials Section 1, 2 for the derivations.

\subsection*{Device fabrication}

To fabricate the desired structure, an 8 nm thick titanium adhesion, a 70 nm thick Au film and a subsequent 5 nm thick titanium layer were successively sequentially deposited onto a Si substrate using electron beam evaporation (EBE; DETECH, DE500C) at a deposition rate of 0.6Å/s under a chamber pressure of 9×10\textsuperscript{-7}Torr. A 545 nm SiO\textsubscript{2} layer followed by a 157 nm Si layer were deposited using plasma-enhanced chemical vapor deposition (PECVD). Next, the PMMA resist (950 A4 from KayakuAM) was spin-coated onto the sample at 3000 rpm for 60s, and baked at 180℃ for 90s, producing a uniform 220nm thick resist film. The reverse designed structure patterns were then exposed onto the PMMA layer using a 50KV electron beam lithography (EBL) system (Raith Voyager), and a subsequent development of the exposed resist was performed to obtain the desired PMMA pattern. A post development bake was carried out on a hot plate at 100°C for 90 s. The final Si structures were obtained via inductively coupled plasma-enhanced reactive ion etching (ICP-RIE, Oxford Instruments, Plasma Pro 80 Cobra). Finally, the sample was immersed in acetone for 5 minutes to completely remove the residual PMMA, followed by ethanol rinsing to ensure a clean and contaminant-free surface.

 The morphology of the metasurface sample was characterized by a scanning electron microscope (Sigma 500, Zeiss).

\subsection*{Numerical Simulation}

The eigenfrequencies of the structure were computed using the Electromagnetic Waves, Frequency Domain (EWFD) module in COMSOL Multiphysics.

The s-polarized reflectance spectrum $R_s$ was obtained through full-wave simulations conducted with the Rigorous Coupled-Wave Analysis (RCWA) module in ANSYS Lumerical FDTD. Both solvers used the same complex refractive indices to capture the influence of material absorption on conversion efficiency: $n_{\rm Si}=3.58 + i\,10^{-4},n_{\rm SiO_2}=1.45 + i\,10^{-4}$. The chosen imaginary part \((k=10^{-4})\) reflects realistic losses in silicon and silicon dioxide films grown by CVD and is critical for accurate prediction of cross‑polarization conversion efficiencies. The experimentally observed shift of the BIC resonance from approximately 1550 nm to 1520 nm is attributed to the slightly higher refractive index \(n_{\rm Si}\) used in the simulations compared to the experimental material, as confirmed by numerical comparison detailed in the Supplementary Materials Section 7.

\subsection*{Data Availability}
The data that support the findings of this study are available from the
corresponding author upon reasonable request.

\bibliography{ref.bib}
\end{document}